\documentclass[aps,preprint,floatfix,nofootinbib,showpacs]{revtex4-1}
\usepackage{graphicx,color,epstopdf,amsmath}

\newcommand{\lm}{{\lambda}}
\newcommand{\pb}{{\;{\rm pb}}}

\begin{document}
\title{
The Golden Mode for a Baryonic $Z'$ Boson at Hadronic Colliders:
$pp/p\bar{p}\to WZ'\to \ell\nu b\bar{b}$
}
\author{ Kingman Cheung$^{1,2}$, Jeonghyeon Song$^{1}$}
\affiliation{
$^1$Division of Quantum Phases \& Devices, School of Physics, 
Konkuk University, Seoul 143-701, Republic of Korea \\
$^2$Department of Physics, National Tsing Hua University, 
Hsinchu 300, Taiwan
}

\date{\today}

\begin{abstract}
  Associated production of a baryonic $Z'$ boson with the $W$ boson
  can account for the excess in $Wjj$ production observed by the CDF
  collaboration at the Tevatron.  We analyze other possible channels
  of this $Z'$ at the Tevatron and at the LHC, including $\gamma Z'$
  and $Z Z'$ with the $Z' \to jj$. We show that the chances of
  confirming this baryonic $Z'$ is better at the Tevatron than at the
  LHC because of the faster growing backgrounds at the LHC.
  Unfortunately the current systematic uncertainties of the order of
  10\% cannot yield any significant excess in both $\gamma Z'$ and $Z
  Z'$ channels at the Tevatron and also at the LHC.  Nevertheless the
  search using the $b\bar b$ decay mode of $Z'$ is much more feasible
  at the LHC, provided that the branching ratio $B(Z' \to b\bar b) >
  0.1$.  In particular, the $W Z' \to \ell\nu b\bar b$ mode has a
  signal-to-background ratio larger than 1. Even with 1 fb$^{-1}$
  luminosity at the LHC it can lead to a high significance level. 
  The $WZ' \to \ell\nu b\bar b$ and $\gamma Z' \to \gamma b
  \bar b$ are also highly observable at the Tevatron.
\end{abstract}

\pacs{}
\maketitle

\section{Introduction}

The recent experimental anomaly presented by the CDF collaboration
was a 3.2$\,\sigma$ excess in the invariant-mass
window $M_{jj} \sim 120 - 160$ GeV 
of the dijet system of the associated production
of a $W $ boson \cite{cdf-wjj}.  
We shall denote it by the CDF $Wjj$ anomaly.
The excess appears to be
a resonance, but the current resolution \cite{cdf-wjj}
cannot tell whether it is a narrow resonance or not.
From the distribution we can naively see that the width
of the resonance appears to be slightly wider than the SM $Z$ boson.
There was a public talk \cite{punzi} very recently that analyzed
a larger data set of 7.3 fb$^{-1}$.  The significance of the peak
becomes more prominent and is at about $4.1\sigma$ significance level.
The extension of the resonance peak shifts slightly to $130-170$ GeV.  
In the rest of this work, we estimate the SM background in this new
invariant-mass window $130 \;{\rm GeV} < M_{jj} < 170\;{\rm GeV}$
and also take $M_{Z'} = 150$ GeV for 
illustration\footnote{The 
D\O\ collaboration \cite{d0} just announced their result
and found no evidence of such a resonance.}.
The CDF anomaly has stimulated a lot of phenomenological activities,
where explanations can be divided into additional gauge bosons 
and variations \cite{our,zprime,joa},
scalar bosons \cite{scalar}, others \cite{others}, 
and within the SM \cite{sm}. 

In a recent work \cite{our}, we proposed a baryonic $Z'$ model to 
explain the anomaly.  The reason for being baryonic is that 
if this $Z'$ has a small leptonic branching ratio,
even $O(1)\%$, it would suffer from strong constraints of the Tevatron
$Z'$ search in the dilepton mode \cite{cdf-z'}.
In addition, its mixing with the SM $Z$ boson
should be extremely small to be compatible with the LEP
electroweak precision data \cite{z'-lep}.
The baryonic $Z'$ model was proposed by Barger, Cheung, and Langacker 
in 1996 \cite{bcl} 
in light of the $R_{b}-R_{c}$ crisis of the LEP precision measurements
at that time \cite{lep}.
Theoretically 
many extensions of the SM have extra $U(1)$ gauge symmetries, 
and thus
additional $Z'$ bosons \cite{paul}.
An interesting possibility is a baryonic $Z'$
from a gauge
symmetry generated by the baryon number \cite{carone}. 
Another possibility is kinetic mixing of the two
$U(1)$'s \cite{babu} to suppress the leptonic couplings.
Here we assume that the model can be embedded in an anomaly-free theory. 

Since this baryonic $Z'$ only couples with quarks,
strong implications at a hadron collider is expected \cite{bcl}
such as
a single $Z'$ production via $s$-channel and
the pair production
processes $(W,Z,\gamma)Z'$ with $Z'\to jj$
with invariant mass $M_{jj}$ peaked at $M_{Z'}$.
The $s$-channel $Z'$ production is buried under the QCD background.
On the contrary
the associated production with a $W$ boson has a good chance to appear,
which can explain the CDF $Wjj$ anomaly \cite{cdf-wjj}.
 
A natural follow-up question is what other signatures
are expected at hadron colliders.
Also, another question is whether the LHC with higher energies
and more luminosities than the Tevatron is 
favorable to probe this baryonic $Z'$ model.
The first clean signature could be the excess in the
$Z jj$ and  $\gamma jj$ channels.  
We show through detail analysis that
it is hard to see the excess in both $\gamma Z' \to \gamma jj$ and 
$Z Z' \to \ell^+ \ell^- jj/\nu\bar{\nu}jj$ channels at the Tevatron
under the current level of systematic uncertainties and jet cuts.
However, with tightened jet cuts and improved systematic uncertainties
it could be promising to test the model in these two channels.
We give details and show how to suppress the backgrounds.
We also analyze the corresponding $Wjj$, $Zjj$, and $\gamma jj$ signals
and backgrounds at the LHC with center-of-mass energies of $7, 10$, and 
14 TeV. However, we found that these channels at the LHC
are not as good as at the Tevatron, because the backgrounds are growing
much faster with energy than the signals.
Nevertheless, we investigate the feasibility of using 
$Z' \to b \bar b$ mode, provided that the branching ratio 
$B(Z' \to b\bar b) > 0.1$.  
We found that the search using $b\bar b$
mode would be much more promising than just the regular jets. 

The organization is as follows. We briefly describe the model in 
Sec. II.  We show in Sec. III the analysis on the $Z jj$ and $\gamma jj$ 
channels at the Tevatron, and in Sec. IV on $Wjj$, $Zjj$, and $\gamma jj$
at the LHC. We study $\gamma b\bar b$, $Z b\bar b$, and $W b\bar b$ 
at the LHC and at the Tevatron in Sec. V. 
We summarize in Sec. VI.

\section{Baryonic $Z'$ model}

Following Ref.~\cite{lang-luo}, 
we consider the Lagrangian describing the
neutral current gauge interactions of the
standard electroweak $SU(2)\times U(1)$ and extra $U(1)$'s,
given by
\begin{equation}
- {\cal L}_{\rm NC} = e J_{\rm em}^\mu A_\mu + \sum_{\alpha=1}^{n}
g_\alpha J^\mu_\alpha Z^0_{\alpha \mu}\;,
\end{equation}
where $Z^0_1$ is the SM $Z$ boson and 
$Z^0_\alpha$ with $\alpha\ge 2$ are the
extra $Z$ bosons in the weak-eigenstate basis.
For the present work we only consider one extra $Z_2^0$ 
mixed with the SM $Z^0_1$ boson. 
The coupling constant $g_1$ is the SM coupling 
$g/\cos\theta_{\rm w}$ where $\theta_{\rm w}$ is the weak mixing
angle. 
In grand unified theories $g_2/g_1$ is 
\begin{equation}
\frac{g_2}{g_1} = \left(\frac{5}{3}\, 
\sin^2\theta_{\rm w}\lambda\right)^{1/2} \simeq
0.62\lambda^{1/2} \,,
\label{eq:g2/g1}
\end{equation}
where the factor $\lambda \sim O(1)$ 
depends on the symmetry breaking pattern and the fermion
sector of the theory.

Since we only consider one additional $Z_2^0$ we can rewrite
the Lagrangian with only the $Z^0_1$ and $Z^0_2$ interactions:
\begin{equation}
-{\cal L}_{Z^0_1 Z^0_2} = 
\frac{g_1}{2}
 Z^0_{1\mu} \left[  \sum_i
 \bar \psi_i \gamma^\mu (g_v^{i(1)} - g_a^{i(1)} \gamma^5 ) \psi_i \right] +
 \frac{g_2}{2}
 Z^0_{2\mu} \left[ \sum_i
 \bar \psi_i \gamma^\mu (g_v^{i(2)} - g_a^{i(2)} \gamma^5 ) \psi_i \right]\;.
\end{equation}
For both SM quarks and leptons
\begin{equation}
g_v^{i(1)} = T_{3L}^i - 2 x_{\rm w} Q_i \,, \qquad
g_a^{i(1)} = T_{3L}^i\,,
\end{equation}
where $T_{3L}^i$ and $Q_i$ are, respectively, the third component of the weak
isospin and the electric charge of the fermion $i$.
We consider the case in which $Z^0_2$ couples only to quarks:
\begin{equation}
g_v^{q(2)} = \epsilon_V \,,\qquad
g_a^{q(2)} = \epsilon_A \,,\qquad
g_v^{\ell(2)} = g_a^{\ell(2)} = 0\;.
\end{equation}
The parameters $\epsilon_V$ and $\epsilon_A$ are the vector 
and axial-vector couplings of $Z^0_2$.
Without loss of generality we choose $\epsilon_V=\sin\gamma$ and  
$\epsilon_A=\cos\gamma$ such that $(\epsilon_V^2 + \epsilon_A^2) $ is
normalized to unity:
\begin{eqnarray}
\left( g_v^{q(2)} \right)^2 +
\left( g_a^{q(2)} \right)^2 = 
\epsilon_V^2 + \epsilon_A^2 = 1.
\end{eqnarray}

The mixing of the weak eigenstates $Z^0_1$ and $Z^0_2$ to form  mass 
eigenstates $Z$ and $Z'$  are parametrized
by a mixing angle $\theta$:
\begin{equation}
\label{mixing}
\left ( \begin{array}{c} Z \\
                         Z'
        \end{array} \right ) = \left( \begin{array}{rr}
                                  \cos\theta & \sin\theta \\
                                 -\sin\theta & \cos\theta
                                      \end{array} \right ) \;
    \left( \begin{array}{c} Z^0_1 \\
                            Z^0_2
            \end{array} \right ) \;.
\end{equation}
%

After substituting  the interactions of the mass eigenstates $Z$ and  
$Z'$ with fermions are 
\begin{equation}
\label{rule}
-{\cal L}_{Z Z' } = \sum_i \frac{g_1}{2} \biggl [
  Z_{\mu} \bar \psi_i \gamma^\mu (v_i - a_i \gamma^5 ) \psi_i  +
  Z'_{\mu} \bar \psi_i \gamma^\mu (v_i' - a_i' \gamma^5 ) \psi_i  \biggr ]\,,
\end{equation}
where
\begin{eqnarray}
v_i = g_v^{i(1)} + \frac{g_2}{g_1} \, \theta \, g_v^{i(2)} \,, &\qquad&
a_i = g_a^{i(1)} + \frac{g_2}{g_1} \, \theta \, g_a^{i(2)} \,, \\
v_i' = \frac{g_2}{g_1}\, g_v^{i(2)} - \theta \, g_v^{i(1)}\,, &\qquad&
a_i' = \frac{g_2}{g_1}\, g_a^{i(2)} - \theta \, g_a^{i(1)}\,.
\end{eqnarray}
Here we have used the valid approximation $\cos\theta\approx 1$ 
and $\sin\theta \approx \theta$.  In the following, we ignore the mixing 
($\theta =0$) such that the precision measurements for the SM $Z$ boson are not
affected, unless stated otherwise. 
For later discussions,
we also express couplings of the $Z'$ boson as
\begin{equation}
 -{\cal L}_{Z'} = g_2 Z'_{\mu} \sum_i \;
  \bar \psi_i \gamma^\mu (  g_{iL}'  P_L + g_{iR}' P_R ) \psi_i
\end{equation}
where the left- and right-handed couplings are given by
$g'_{iL,iR} = (g_v^{i(2)}\pm g_a^{i(2)}) /2$
in the limit of no $Z$-$Z'$ mixing.

The decay width of
$Z' \to q \bar q$ is given by
\begin{equation}
\label{width}
\Gamma (Z' \to q\bar q ) = \frac{G_F M_{Z}^2 }{6\pi \sqrt{2} }
N_c  C(M_{Z'}^2) M_{Z'} \sqrt{ 1 - 4x} \left[
\left(v_{q}^{\prime}\right)^2 (1+2x) +  
\left(a_{q}^{\prime}\right)^2 (1-4x) \right] \,,
\end{equation}
where $G_F$ is the Fermi coupling constant,
$C(M_{Z'}^2) = 1+\alpha_s/\pi + 1.409 (\alpha_s/\pi)^2 -12.77 (\alpha_s/\pi)^3$, 
$\alpha_s = \alpha_s (M_{Z'})$ is the strong coupling at the scale $M_{Z'}$,
$x=m_f^2/M_{Z'}^2$, 
and $N_c=3$ or 1 if $f$ is a quark or a lepton,
respectively.
The $Z'$ width is proportional to $\lambda$, which sets the strength of
the $Z'$ coupling. For $\lambda=1$ the total $Z'$ width is
\begin{equation}
\frac{\Gamma_{Z'} }{ M_{Z'}}
 = 0.022 \quad {\rm for}\ M_{Z'} < 2m_t  \;.
\end{equation}
The width would be increased somewhat if there are open channels for decay
into the top quark, superpartners, and other exotic particles. 
Essentially, it is a narrow resonance.


\section{UA2 Constraint and  fit to the CDF $Wjj$ excess}

The dominant production of the $Z'$ boson at a hadron collider 
is through the $q\bar
q\to Z'$ subprocess with the cross section in the narrow $Z'$ width
approximation of \cite{collider}
\begin{equation}
\label{eq:qqZp}
\hat\sigma(q\bar q\to Z') = K \frac{2\pi}{3} \frac{G_F\,M_{Z}^2}{\sqrt2}
\left[ \left(v_q'\right)^2 + \left(a_q'\right)^2 \right]
 \delta\! \left(\hat s - M_{Z'}^2\right) \,.
\end{equation}
The $K$-factor represents the enhancement from higher order QCD processes,
estimated to be 
$K = 1 + \frac{\alpha_s(M_{Z'}^2)}{2\pi} \frac{4}{3} \left( 1 +
\frac{4}{3}\pi^2 \right) \simeq 1.3$ \cite{collider}.
When the mixing is ignored, we have
\begin{equation}
\left(v_q'\right)^2 + \left(a_q'\right)^2 = (0.62)^2 \lambda
\end{equation}
and the cross section is independent of the parameter $\gamma$ 
as long as  $\epsilon_V^2 + \epsilon_A^2 = 1$.

Note that all the current and previous dijet-mass searches 
\cite{cdf-dijet} at the Tevatron
are limited to $M_{jj}> 200$ GeV, which are not applicable to the present
$Z'$ with $M_{Z'} \approx 145-150$ GeV.
The relevant dijet data were from UA2 collaboration
with collision 
energy at $\sqrt{s} =630$ GeV.  
The UA2 collaboration detected 
the $W+Z$ signal in the dijet mass
region $48<M_{jj}<138$~GeV and 
put upper bounds on $\sigma B(Z'\to jj)$
over the range $80<M_{jj}<320$~GeV~\cite{ua2}. 
The analysis on the UA2 data were shown 
in Fig.~1 of Ref.~\cite{bcl}: 
the allowed values are $\lambda \alt 1$ for
$M_{Z'} = 100 - 180$ GeV, given the uncertainty in the $K$-factor in 
the theoretical cross section calculation
and the difficulty in obtaining an experimental bound by
subtraction of a smooth background.
Under the assumption that $Z'$ coupling to the
up-type quark is the same as that 
to the down-type quark,
the UA2 constraint is therefore given by
\begin{eqnarray}
\hbox{ UA2 constraint: }
  \lambda < 1.
\label{ua2}
\end{eqnarray}

Our phenomenological model does not specify
each $Z'$ coupling,
$g'_{iL}$ and $g'_{iR}$.
The single production of $Z'$ in Eq.~(\ref{eq:qqZp})
does not depend on the relative size of $g'_{iL}$ 
to $g'_{iR}$ as long as the normalization of
$\left( g'_{iL} \right)^2 + \left( g'_{iR} \right)^2=1/2$
holds.
Note that the production of $\gamma Z'$ is also proportional
to $\left( g'_{iL} \right)^2 + \left( g'_{iR} \right)^2$
because of the vector coupling of the photon with fermions.
On the contrary, the production of $WZ'$,
which may explain the CDF $Wjj$ anomaly,
only
depends on the left-handed coupling because of the presence of the
$W$ boson.

In Ref.~\cite{our}, we have assumed the democratic 
coupling of $Z'$, 
and chosen $\epsilon_V = \epsilon_A = 1/\sqrt{2}$, or equivalently
\begin{equation}
\label{coup}
g_{iL}' = \frac{1}{\sqrt{2}},\qquad  g_{iR}' = 0 \;. 
\end{equation}
Note that the purely left-handed coupling for the $Z'$ boson
maximizes $\sigma(WZ' \to W jj)$ with a given value of $\lm$.
Together with the  choice of
$\lm=1$, 
which is maximally allowed by the UA2 data,
we could explain the cross section 
of $\sigma(WZ') = 4$ pb 
claimed by CDF \cite{cdf-wjj}.
The relative size of $g_{iL}'/ g_{iR}'$ can have some other combinations
if the excess in the CDF
data is estimated differently, \textit{e.g.},
$2.5-4$ pb as suggested in some of the papers~\cite{zprime}.
If we set
$\sigma(WZ') = 3$ pb and $\lm=1$,
we have $g^{i(2)}_L =
 \frac{1}{\sqrt{2}}\frac{\sqrt{3}}{2}$
and
$g^{i(2)}_R = \frac{1}{\sqrt{2}}\frac{1}{2}$.
Finally the other channel $Z Z'$, which was shown in 
Ref.~\cite{joa}, depends dominantly on the left-handed coupling:
at the LHC this dominance becomes more significant. 
In order to maximize
the production of $ZZ'$ and $WZ'$ under the UA2 constraint,
the choice of couplings is now clear
to be in Eq.~(\ref{coup}).
In the rest of the paper, we shall stick to this choice.

\section{$WZ'$, $ZZ'$ and $\gamma Z' $  Production at the Tevatron}

The associated production of $Z'$ with a $W$ boson goes 
through the
$t$- and $u$-channel exchange of quarks.
The $s$-channel $W$ boson
exchange is highly suppressed because of the negligible 
mixing angle between the SM $Z$ boson and the $Z'$.
Consequently, we expect similar or even larger cross sections for
$M_{Z'} \sim M_Z$ than the SM $WZ$ production in which there is a
delicate gauge cancellation among the $t$-, $u$-, and $s$-channel
diagrams.  
The cross section at the Tevatron energy $\sqrt s =
1.96$~TeV is about 4 pb
for $\lambda=1$ and the normalized $Z'$ coupling in Eq.~(\ref{coup}).  We have
included a $K$-factor of $K=1.3$ 
to approximate next-to-leading order
QCD contributions \cite{ohnemus}.


As shown in Ref.~\cite{bcl} other associated production channels,
$\gamma Z',\; Z Z'$, and $Z' Z'$ are possibly observable, provided
that the current excess is due to $WZ'$ production.
With the same parameters, we have 
$\sigma(ZZ') \simeq 1.3 \pb$, 
$\sigma(\gamma Z') \simeq 0.7 \pb$,
and $\sigma(Z'Z') \simeq 0.4 \pb$
for $M_{Z'}=145$ GeV.
For the
acceptance
on the final state photon,
we have imposed $p_T(\gamma)>50 \;{\rm GeV}$
and $|\eta(\gamma)|<1.1$ \cite{photon:cut}.
Simply from the signal cross sections, one may tempt to conclude that the 
$WZ'$ channel is the most likely one to be observed, followed by 
$ZZ'$ and $\gamma Z'$. 
However, one cannot easily draw this conclusion
without working out the corresponding backgrounds. The same is also
true to the LHC.  In fact, we shall show in the next section
that the backgrounds grow faster than the signals such that the situation
at the LHC is no better than that at the Tevatron.

The irreducible backgrounds to the $(\gamma,W,Z)Z'$ signals with $Z'\to  jj$
arise from the $(\gamma,W,Z) jj $ final states.
We calculate the backgrounds using the MADGRAPH package \cite{madgraph}.
It was mentioned in Refs.~\cite{cdf-wjj,cdf-Z} and in Ref.~\cite{cdf-gamma}
that no significant excess is observed in $Zjj$ and $\gamma jj$ channel,
respectively. 
We shall show that with current systematic uncertainties
of level 10\% \cite{cdf-wjj,cdf-Z,cdf-gamma,thesis} 
and a similar set of
jet cuts, no significant excess can be observed in both channels.
Some improvements are possible if we further tighten the cuts on the jets.
Details of the signal and background cross sections
are summarized in Table~\ref{table-teva}.
Here we have used an integrated luminosity of 10 fb$^{-1}$
in deriving the significance.

The initial choice for jet cuts is the same as the jet cuts in 
Ref.~\cite{cdf-wjj}:
\begin{eqnarray}
&& E_{Tj} > 30 \;{\rm GeV}, \quad |\eta_j| < 2.4, \quad p_{Tjj} > 40\;{\rm GeV}\;,
  \nonumber  \\
&& 130\;{\rm GeV} < M_{jj} < 170\; {\rm GeV} \label{jetcut1} \;.
\end{eqnarray}
Leptonic cuts for $Z \to \ell^+ \ell^-$ are
\begin{equation}
\label{leptoncut}
p_{T\ell} > 25\;{\rm GeV},\quad |\eta_\ell| < 2.8 \;,
\end{equation}
and the photon cuts are
\begin{equation}
p_T(\gamma)>50 \;{\rm GeV}\,, \quad |\eta(\gamma)|<1.1 \label{photon}\;.
\end{equation}

In order to understand the lack of the excess in the
$\gamma jj$ and $Z jj$ yet,
we first discuss these two channels with data 
for ${\cal L}=4.3\,
{\rm fb}^{-1}$.
For the $Zjj$ channel
we have 
$\sigma_{\rm signal}: \sigma_{\rm bkgd} = 26 \,{\rm fb}: 150\,{\rm fb}$. 
It would give a significance of
$S/(\sqrt{B} \oplus 0.1 B) \approx 1.5\sigma$, 
where the factor $0.1$ is the systematic uncertainties. 
With the same set of jet cuts and photon cuts in Eq.~(\ref{photon}) 
to the $\gamma jj$ channel,
we obtain $\sigma_{\rm signal}: \sigma_{\rm bkgd} = 0.5 \,{\rm pb}: 8.8\,{\rm pb}$,
which gives a significance of 
$S/(\sqrt{B} \oplus 0.1 B) \approx 0.5\sigma$.
Therefore, we cannot observe any significant excess 
in both channels at the Tevatron,
in accord with the claims in Refs.~\cite{cdf-wjj,cdf-gamma}.

Nevertheless, if we tighten the jet cuts the backgrounds will suffer 
more than the signals. With 
\begin{eqnarray}
&& E_{Tj} > 50\,{\rm GeV}, 
\quad |\eta_j| < 2.4, \quad p_{Tjj} > 40\;{\rm GeV}\;,
  \nonumber  \\
&& 130\;{\rm GeV} < M_{jj} < 170\; {\rm GeV} \label{jetcut2}
\end{eqnarray}
and 
${\cal L}=10\, {\rm fb}^{-1}$, the significance can improve to 
$2.4\sigma$ and $1\sigma$ for $Zjj$ and $\gamma jj$ channel, respectively.
If the systematic uncertainties can be reduced to an ideal level of $2-3$\%
the significance can be further improved to $5\sigma$ and $4\sigma$,
respectively.  Details of the signal and background cross sections
are summarized in Table~\ref{table-teva}.

\begin{table}[tb!]
\centering
\caption{\small \label{table-teva}
Signal and background cross sections for $\gamma Z' \to \gamma jj$,
$Z Z' \to \ell^+ \ell^- jj$, $Z Z' \to \nu \bar\nu  jj$,
and $W Z' \to \ell \nu jj$ at the Tevatron for $M_{Z'} = 150$ GeV, 
with jet cuts defined in Eqs.~(\ref{jetcut1}) and (\ref{jetcut2}). 
Photons cuts for $\gamma jj$ are in Eq.~(\ref{photon}), the lepton 
cuts for $Zjj \to \ell^+ \ell^- jj $ are in Eq.~(\ref{leptoncut}), 
and the leptonic cuts for $W jj \to \ell \nu jj$ are 
$p_{T\ell} > 20$ GeV, $|\eta_\ell| < 1$, and $\not\!{E}_T > 25$ GeV.
Significance levels are calculated with 10\%, 6\%, and 2\% systematic 
uncertainty, shown in the last three columns, respectively, 
with ${\cal L}=10\;{\rm fb}^{-1}$, 
}
\begin{ruledtabular}
\begin{tabular}{lccccc}
Jet cuts (Tevatron) & $\sigma_{\rm signal}$ &  $\sigma_{\rm bkgd}$  
   & $\frac{S}{\sqrt{B} \oplus 0.1 B}$ & $\frac{S}{\sqrt{B} \oplus 0.06 B}$
   & $\frac{S}{\sqrt{B} \oplus 0.02B}$ \\
\hline 
& \multicolumn{5}{l}{$\gamma Z' \to \gamma jj$} \\
$E_{Tj} > 30\; {\rm GeV}$ & 0.49 pb & 8.79 pb & 0.55 & 0.92 & 2.7 \\
$E_{Tj} > 50\; {\rm GeV}$ & 0.28 pb & 3.05 pb & 0.91 & 1.5 & 4.4 \\
\hline
  & \multicolumn{5}{l}{$Z Z' \to \ell^+ \ell^- jj \; (\ell=e,\mu)$} \\
$E_{Tj} > 30\; {\rm GeV}$ & 0.026 pb & 0.15 pb & 1.7 & 2.6 & 5.3 \\
$E_{Tj} > 50\; {\rm GeV}$ & 0.015 pb & 0.057 pb & 2.4 & 3.6 & 5.7 \\
\hline
  & \multicolumn{5}{l}{$Z Z' \to \nu \bar\nu jj \; 
(\nu =\nu_e,\nu_\mu,\nu_\tau)$ with $\not\!{E}_T > 40$ GeV} \\
$E_{Tj} > 30\; {\rm GeV}$ & 0.12 pb & 0.80 pb & 1.5 & 2.5 & 6.8 \\
$E_{Tj} > 50\; {\rm GeV}$ & 0.072 pb & 0.30 pb & 2.4 & 3.8 & 8.8 \\
\hline
  & \multicolumn{5}{l}{$W Z' \to \ell \nu  jj \; (\ell =e,\mu)$} \\
$E_{Tj} > 30\; {\rm GeV}$ & 0.16 pb & 0.92 pb & 1.8 & 2.9 (5\%: 3.5) & 7.9\\
$E_{Tj} > 50\; {\rm GeV}$ & 0.096 pb & 0.35 pb & 2.7 & 4.4 (5\%: 5.2) & 10.4 
\end{tabular}
\end{ruledtabular}
\end{table}

In Table~\ref{table-teva}, we also include the channel 
$Z Z' \to \nu \bar \nu jj$, where we have imposed a transverse missing 
energy cut $\not\!{E}_T > 40$ GeV.  Though it may have larger background,
it enjoys a larger branching ratio of $B(Z \to \nu \bar \nu)$. 
Experimentally,
one can combine both the leptonic and invisible modes in the search to
increase the event rates.

Note that our method to estimate the significance is
very basic simply by taking the systematic and statistical uncertainties
as independent quantities and adding them in quadrature. The information 
on the current systematic uncertainties were gathered in a number of 
CDF papers \cite{cdf-wjj,cdf-dijet,cdf-gamma} and the thesis 
in Ref.~\cite{thesis}. The dijet systematic uncertainties in the
signal region are all about 10\%. The significance presented here
can only be compared to one another in the relative sense. The one
presented in the CDF paper \cite{cdf-wjj} was based on the true
data and through detailed background studies. 
In order to achieve a significance of about $3\sigma$ in the $Wjj$ channel,
we need a systematic uncertainty of $5-6\%$ 
in our basic quadrature method,
shown in the  second last
row of Table~\ref{table-teva}.

Brief comments on the dominance of the systematic uncertainties
are in order here.
With about a 1 pb cross section for a total
10 fb$^{-1}$ luminosity,
we have about $10^4$ events.
The 10\% systematic uncertainty ($\sim 10^3$ events)
is much larger than the statistical uncertainty ($\sim 10^2$ events).
Unless there is significant improvement 
of systematic uncertainty, a
larger data set does not help to improve our signal
since the significance in this case is mainly
proportional to $S/B$.
However, we expect that
a larger data set would help us understand the background better,
and thus reduce the systematic uncertainties.

\section{$Wjj$, $Zjj$, and $\gamma jj$ channels at the LHC}

For the SM backgrounds at the LHC,
the jet cuts are chosen as
\begin{eqnarray}
&& E_{Tj} > 50\,{\rm GeV}, \quad 
|\eta_j| < 2.5, \quad p_{Tjj} > 40\;{\rm GeV}\;,
  \nonumber  \\
&& 130\;{\rm GeV} < M_{jj} < 170\; {\rm GeV} \label{jetcut-lhc} \;.
\end{eqnarray}
The cuts on the final-state photon are
\begin{equation}
\label{photon-lhc}
p_T(\gamma)>50 \;{\rm GeV},\;\;\; |\eta(\gamma)|< 2.0 .
\end{equation}
The cuts on the final-state charged leptons (electrons and muons) are
\begin{equation}
\label{lepton-lhc}
p_{T\ell} > 25\;{\rm GeV},\quad |\eta_\ell| < 2.5 \;.
\end{equation}
We use the same leptonic cuts for the $Z \to \ell^+ \ell^-$ and
$W^\pm \to \ell^\pm \nu_\ell$, and additional energy missing cut in 
the $W$ decay:
\begin{equation}
\label{ptmiss}
\not\!{E}_{T} > 40\;{\rm GeV}\;.
\end{equation}
We show the cross sections and significance in Table~\ref{table-lhc}. 
One can see that the signal-to-background ratios deteriorate when
the energy of collisions increases, simply because the backgrounds
outgrow the signal rapidly.  The significance levels that we can achieve
are not as good as those at the Tevatron.  Unless the systematic uncertainties
can be controlled to the level of 2\%, 
the observability at the
LHC would be hard. 

\begin{table}[th!]
\centering
\caption{\small \label{table-lhc}
Signal and background cross sections for $\gamma Z' \to \gamma jj$,
$Z Z' \to \ell^+ \ell^- jj$, $Z Z' \to \nu \bar\nu  jj$,
and $W Z' \to \ell \nu jj$ at the LHC (7, 10, 14 TeV) for $M_{Z'} = 150$ GeV, 
with jet cuts defined in Eqs.~(\ref{jetcut-lhc}). 
Photons cuts for $\gamma jj$ are in Eq.~(\ref{photon-lhc}), 
the lepton cuts 
are in Eq.~(\ref{lepton-lhc}) and
in Eq.~(\ref{ptmiss}).
Significance levels are calculated with 10\%, 6\%, and 2\% 
systematic uncertainty, shown in the last three columns, respectively,
with the corresponding luminosity ${\cal L}$.
}
\begin{ruledtabular}
\begin{tabular}{lccccc}
  LHC  &  $\sigma_{\rm signal}$ &  $\sigma_{\rm bkgd}$  
   & $\frac{S}{\sqrt{B} \oplus 0.1 B}$ & $\frac{S}{\sqrt{B} \oplus 0.06 B}$ &
   $\frac{S}{\sqrt{B} \oplus 0.02B}$ \\
\hline 
&      \multicolumn{5}{l}{$\gamma Z' \to \gamma jj$} \\
7 TeV (1 fb$^{-1}$) 
 & 0.83 pb &  40.9 pb & 0.20 &0.34 &0.99  \\
10 TeV (10 fb$^{-1}$)  
 & 1.2 pb & 66.6 pb & 0.18 &0.31 &0.91  \\
14 TeV (100 fb$^{-1}$) 
 & 1.7 pb & 104.2 pb & 0.16 &0.27 &0.82  \\
\hline
 & \multicolumn{5}{l}{$Z Z' \to \ell^+ \ell^- jj \; (\ell=e,\mu)$} \\
7 TeV (1 fb$^{-1}$) 
 & 0.051 pb & 0.72 pb & 0.66& 1.0 &1.7  \\
10 TeV (10 fb$^{-1}$) 
 &0.081 pb & 1.27 pb & 0.64 & 1.1 &2.9  \\
14 TeV (100 fb$^{-1}$) 
  &0.12 pb & 2.08 pb & 0.58 & 0.96 & 2.9 \\
\hline
 & \multicolumn{5}{l}{$Z Z' \to \nu \bar\nu jj \; 
(\nu =\nu_e,\nu_\mu,\nu_\tau)$} \\
7 TeV (1 fb$^{-1}$) 
 &  0.28 pb &  4.92 pb & 0.56& 0.92 &2.3  \\
10 TeV (10 fb$^{-1}$) 
  & 0.46 pb & 9.16 pb & 0.50 &0.84 & 2.5  \\
14 TeV (100 fb$^{-1}$) 
 & 0.70 pb & 15.6 pb & 0.45 &0.75 &2.2  \\
\hline
 & \multicolumn{5}{l}{$W Z' \to \ell \nu  jj \; (\ell =e,\mu)$} \\
7 TeV (1 fb$^{-1}$) 
  & 0.36 pb &  4.75 pb & 0.75& 1.2 &3.1 \\ 
10 TeV (10 fb$^{-1}$) 
  &0.57 pb & 8.45 pb & 0.67& 1.1 &3.3  \\
14 TeV (100 fb$^{-1}$) 
 & 0.84 pb & 13.8pb & 0.61 & 1.0 &3.0 
\end{tabular}
\end{ruledtabular}
\end{table}

\section{Using the $Z' \to b \bar b$ Mode at the LHC and the Tevatron}

Improvement can be made if the $Z'$ boson decays into $b\bar b$,
because both CMS and ATLAS experiments have a decent $B$-tagging
efficiency and a rather low mistag probability \cite{b-tagging}.  
To make the analysis
simple enough but still semi-realistic we choose the $B$-tagging
and mistag efficiencies to be 
\begin{equation}
\label{B}
 \epsilon_b \approx 0.5,\;\;\; \epsilon_{\rm mistag} = 0.01 \;.
\end{equation}
We assume the $b\bar b$ branching ratio of the $Z'$ to be
\begin{equation}
\label{br}
B(Z' \to b \bar b) = 0.15 \;,
\end{equation}
where this choice is close to the democratic choice and at the
same time consistent with the dedicated $\ell\nu b \bar b$ search
\cite{cdf-wh} at the Tevatron.
The cuts on the $b$-quark jets, leptons, and photons are the same as
those presented in the last section.

\begin{table}[th!]
\centering
\caption{\small \label{table-lhc-bb}
Signal and background cross sections for $\gamma Z' \to \gamma b\bar b$,
$Z Z' \to \ell^+ \ell^- b\bar b$, $Z Z' \to \nu \bar\nu  b\bar b$,
and $W Z' \to \ell \nu b\bar b$ at the LHC (7, 10, 14 TeV) for 
$M_{Z'} = 150$ GeV, 
with jet cuts defined in Eqs.~(\ref{jetcut-lhc}). 
Photons cuts for $\gamma b\bar b$ are in Eq.~(\ref{photon-lhc}), the lepton 
cuts for $Zb\bar b \to \ell^+ \ell^- b \bar b$ and $Wb\bar b\to \ell \nu b
\bar b$ channels are in Eq.~(\ref{lepton-lhc}) and
in Eq.~(\ref{ptmiss}).
We have assumed $B$-tagging and mistag efficiencies as in Eq.~(\ref{B})
and branching ratio $B(Z'\to b\bar b) = 0.15$.
Significance levels are
calculated with 10\%, 6\%, and 2\% systematic uncertainty, shown in the last
three columns, respectively.
}
\begin{ruledtabular}
\begin{tabular}{lccccc}
  LHC  &  $\sigma_{\rm signal}$ &  $\sigma_{\rm bkgd}$  
   & $\frac{S}{\sqrt{B} \oplus 0.1 B}$ & $\frac{S}{\sqrt{B} \oplus 0.06 B}$ 
  & $\frac{S}{\sqrt{B} \oplus 0.02B}$ \\
\hline 
&    \multicolumn{5}{l}{$\gamma Z' \to \gamma b\bar b$} \\
7 TeV (1 fb$^{-1}$) 
    & 0.031 pb & 0.083 pb & 2.5 & 3.0 & 3.3  \\
10 TeV (10 fb$^{-1}$) 
   & 0.046 pb & 0.16 pb & 2.8 & 4.5 & 9.1  \\
14 TeV (100 fb$^{-1}$) 
   & 0.064 pb & 0.28 pb & 2.3 &3.8 & 11.0  \\
\hline
 &  \multicolumn{5}{l}{$Z Z' \to \ell^+ \ell^-b\bar b \; (\ell=e,\mu)$} \\
7 TeV (1 fb$^{-1}$) 
   &  0.0019 pb & 0.0031 pb & 1.1& 1.1 & 1.1  \\
10 TeV (10 fb$^{-1}$) 
  & 0.0030 pb & 0.0065 pb & 2.9 & 3.4& 3.7  \\
14 TeV (100 fb$^{-1}$) 
   &  0.0045 pb & 0.012 pb & 3.5& 5.5 &10.4  \\
\hline
 &  \multicolumn{5}{l}{$Z Z' \to \nu \bar\nu b\bar b \; 
(\nu =\nu_e,\nu_\mu,\nu_\tau)$} \\
7 TeV (1 fb$^{-1}$) 
  &  0.011 pb & 0.017 pb & 2.4& 2.5 & 2.5  \\
10 TeV (10 fb$^{-1}$) 
  &  0.017 pb & 0.037 pb & 4.1& 5.9& 8.4  \\
14 TeV (100 fb$^{-1}$) 
  &  0.026 pb & 0.071 pb & 3.7 & 6.0 & 15.9  \\
\hline
 &  \multicolumn{5}{l}{$W Z' \to \ell \nu b\bar b \; (\ell =e,\mu)$} \\
7 TeV (1 fb$^{-1}$) 
  &  0.014  pb & 0.0047 pb & 6.1& 6.2 & 6.2 \\ 
10 TeV (10 fb$^{-1}$) 
   & 0.021 pb & 0.0074 pb & 18.9& 22.1& 24.5  \\
14 TeV (100 fb$^{-1}$) 
  &  0.032 pb & 0.011  pb & 27.4 & 42.6 & 79.1 
\end{tabular}
\end{ruledtabular}
\end{table}

We show the cross sections and significance at the LHC
in Table~\ref{table-lhc-bb}. 
We can see that the backgrounds decrease substantially, 
because 
the regular jets are dominated by gluons and light quarks. When
one demands $b$ quarks, it decreases tremendously. 
On the other hand, the
signal is only down by a factor $B(Z'\to b \bar b)$ here.  Then both
the signal and background are subject to the $B$-tagging efficiency. 
The signal-to-background ratios improve significantly in all channels:
especially $Wb\bar b \to \ell\nu b\bar b$ has the ratio larger than 1.
Even with an easy systematic uncertainty of 10\% and at 7 TeV with only 
1 fb$^{-1}$, the $Wb\bar b$ channel is very feasible. On the other hand,
the $\gamma b\bar b$ channel needs a better systematic uncertainty in order
to be observed, and the $Z b\bar b \to (\ell^+ \ell^- / \nu \bar \nu) b\bar b$
needs more luminosity.
In conclusions, the search using $Z' \to b \bar b$ is far better than regular jets,
provided that $B(Z' \to b \bar b)> 0.1$. 

One may wonder if using $Z' \to b\bar b$ mode at the Tevatron would also
be useful. We repeat the exercise for the Tevatron and list the signal
and background cross sections and significance in Table~\ref{table-tev-bb}.
Note that we require to have double $B$-tags and that is the reason why
the background is reduced so significantly from those shown in 
Table~\ref{table-teva}.  The signal is only reduced by the branching
ratio $B(Z' \to b\bar b)$ and the double $B$-tagging efficiency.
We only show the significance levels with 10\% systematic uncertainty,
because the significance levels are high enough that one should be
able to see the excess.
According to the significance only,
the $WZ' \to \ell \nu b\bar{b}$ with double $B$-tagging
is the most promising channel
to probe this baryonic $Z'$ boson.
Considering the statistically large enough events,
$\gamma Z' \to \gamma b\bar{b}$ is also very favorable.

\begin{table}[tb!]
\centering
\caption{\small \label{table-tev-bb}
Signal and background cross sections for $\gamma Z' \to \gamma b\bar b$,
$Z Z' \to \ell^+ \ell^- b\bar b$, $Z Z' \to \nu \bar\nu  b\bar b$,
and $W Z' \to \ell \nu b\bar b$ at the Tevatron for $M_{Z'} = 150$ GeV, 
with $b$-jet cuts defined in Eqs.~(\ref{jetcut1}) and (\ref{jetcut2}). 
Photons cuts for $\gamma b\bar b$ are in Eq.~(\ref{photon}), the lepton 
cuts for $Zjj \to \ell^+ \ell^- jj $ are in Eq.~(\ref{leptoncut}), 
and the leptonic cuts for $W jj \to \ell \nu jj$ are 
$p_{T\ell} > 20$ GeV, $|\eta_\ell| < 1$, and $\not\!{E}_T > 25$ GeV.
The $B$-tagging and mistag efficiencies are as in Eq.~(\ref{B})
and branching ratio $B(Z'\to b\bar b) = 0.15$.
Significance levels are calculated with 10\% systematic 
uncertainty and ${\cal L}=10\;{\rm fb}^{-1}$.
}
\begin{ruledtabular}
\begin{tabular}{lccc}
$b$-Jet cuts (Tevatron) & $\sigma_{\rm signal}$ &  $\sigma_{\rm bkgd}$  
   & $\frac{S}{\sqrt{B} \oplus 0.1 B}$ \\
\hline 
& \multicolumn{3}{l}{$\gamma Z' \to \gamma b\bar b$} \\
$E_{Tb} > 30\; {\rm GeV}$ & 18.4 fb & 22.5 fb & 6.8   \\
$E_{Tb} > 50\; {\rm GeV}$ & 10.5 fb & 10.3 fb & 7.3   \\
\hline
  & \multicolumn{3}{l}{$Z Z' \to \ell^+ \ell^- b\bar b \; (\ell=e,\mu)$} \\
$E_{Tb} > 30\; {\rm GeV}$ & 0.97 fb & 0.59 fb & 3.9   \\
$E_{Tb} > 50\; {\rm GeV}$ & 0.56 fb & 0.30 fb & 3.2   \\
\hline
  & \multicolumn{3}{l}{$Z Z' \to \nu \bar\nu b\bar b \; 
(\nu =\nu_e,\nu_\mu,\nu_\tau)$ with $\not\!{E}_T > 40$ GeV} \\
$E_{Tb} > 30\; {\rm GeV}$ & 4.5 fb & 4.1 fb & 5.9   \\
$E_{Tb} > 50\; {\rm GeV}$ & 2.7 fb & 1.5 fb & 6.5   \\
\hline
  & \multicolumn{3}{l}{$W Z' \to \ell \nu  b\bar b \; (\ell =e,\mu)$} \\
$E_{Tb} > 30\; {\rm GeV}$ & 6.0 fb & 2.4 fb & 11  \\
$E_{Tb} > 50\; {\rm GeV}$ & 3.6 fb & 1.3 fb & 9.3   
\end{tabular}
\end{ruledtabular}
\end{table}

\section{Conclusions}

In summary, we have shown that a baryonic $Z'$ boson can explain the
excess in the invariant-mass window $130 - 170$ GeV in the 
dijet system of $Wjj$ production.  
Such a $Z'$ boson without 
leptonic couplings is not subject to the current dilepton limits on extra 
gauge bosons.  Yet, the strongest constraint comes from the dijet 
search of the UA2 data, from which the size of coupling, proportional
to $\sqrt{\lambda}$, 
is constrained to be $\lambda \alt 1$.  With $\lambda =1$ we are
able to explain the required cross section of 4 pb in the excess
window. 
We have also shown that it is hard to see the excess in
$\gamma Z' \to \gamma jj$ and $Z Z' \to \ell^+ \ell^- jj$ channels at the Tevatron
under the current systematic uncertainties and jet cuts.
However, with tightened jet cuts and improved systematic uncertainties
it could be promising to test the excess in these two channels.

The situation at the LHC with $\sqrt{s} = 7,10,14$ TeV does not improve
because the backgrounds grow with energy much faster than the signal.
The signal-to-background ratios drop and so do the significance levels.
On the other hand, we have shown that if $B(Z'\to b\bar b) > 0.1$ we can use
the $\gamma b\bar b$, $Zb\bar b$, and $Wb\bar b$ channels and the
observability improves substantially at the LHC.  In particular, 
the $W b\bar b \to \ell\nu b\bar b$ mode has a signal-to-background ratio
larger than 1 and very high significance levels.  We urge the LHC
experimenters to search for the $WZ' \to \ell \nu b\bar b$ as well.

A similar urge is also applicable to the Tevatron. We have shown that the 
$WZ'\to \ell \nu b\bar b$, $\gamma Z' \to \gamma b\bar b$,
and $Z Z' \to \nu \bar \nu b\bar b$ channels are observable with the
present level of uncertainties.

\section*{\bf Acknowledgments}
This research was supported in parts by the NSC under Grant
Nos. 99-2112-M-007-005-MY3 and by WCU program through the NRF funded by 
the MEST (R31-2008-000-10057-0).


\end{document}